\def \rem#1 {{}}
\def \com#1 {{\bf \it #1 }}
\def \p#1 {{\bf #1 \par }}
\def \rej#1 {{}}

\def \endnote#1 {{\small [#1]}}
\def \deg#1 {$#1^\circ$}

\documentstyle[twocolumn,prl,aps,epsf]{revtex}
\begin{document} 
\draft 
\title
{Effective  potentials for 6-coordinated Boron: a structural approach}

\author{W.-J. Zhu and
C. L. Henley}
\address{LASSP, Cornell University, Ithaca NY 14853-2501}

\maketitle

%\widetext

\begin{abstract}

We have built an {\it ab-initio} LDA energy database with over 60
hypothetical extended structures of pure Boron, in each of which the
coordination environment of each atom is equivalent.  Focusing on
eleven 6-coordinated structures, which are most relevant to the
observed Boron phases, we deduce a nearest-neighbor-only 2-body
interaction, refined by potentials that depend on angles and asymmetry
within the local atomic environment.  The resulting effective
potentials describe the 6-coordinated structures with an average error
of 0.1 eV/atom, and favor the environment most often seen in real
Boron.

\end{abstract}

\pacs{	PACS. 34.20.Cf,  61.50.Lt, 71.20.Mq, \rem{61.66.Bi,} 33.15.Dj }

\narrowtext  %<-to get two columns a la PRB

\rem{section INTRODUCTION}
\label{sec:intro}

Boron as the prime candidate for a covalent quasicrystal has inspired
 activities in both experimental search and theoretical assessment,
 leading to identification of a new phase \cite{kimura} and interests
 in nanotubes \cite{alex}.  Accurate and fast structural energy
 estimators, capable of handling complex extended structures, can
 facilitate further investigations of these proposed structures,
 (through Monte-Carlo or Molecular Dynamics exploration), and might
 elucidate the structures of the unsolved or amorphous phases.  While
 {\it ab-initio} methods have been used for comparing specific Boron
 structures\cite{kleinman,mailhoit}, we seek a less computationally
 intensive effective theory, in the style of the environment-dependent
 effective interatomic potentials, which have successfully described
 Silicon bonding \cite{tersoff,bazant} but are not yet available for
 Boron.
  
The bonding behavior of Boron is unique among the elements and allows
a number of complex structures.  The atomic structures have been
refined for only the metastable $\alpha_{12}, T_{50}$ and the stable
$\beta_{105}$ \cite{solvedphase}, while various complex phases remain
unsolved \cite{donohue}.  Icosahedral clusters of various scales are
present in all the known atomic arrangements, leading to unusual
coordination environments and geometric patterns \cite{hoard}, quite
unlike those in the well-studied covalent (e.g. Silicon) systems.  In
the case of the structurally simplest $\alpha_{12}$ phase, accurate
electronic band structure \cite{kleinman}, lattice dynamics studies
and force models \cite{shirai} all confirm the presence of 3-centered
bonds, and the coexistence of soft metallic and strong covalent bonds
both within and between icosahedra \cite{longuett}.

To design a classical potential (total energy function) that describes
not only the uncommon bonding mechanisms and large-scale structural
features, but also the unforeseen local configurations that may arise
in quasicrystal models, we assume a local form $ E_{total}
=\sum_{i}E_i$, where $E_i$, the site energy of each atom, depends on
its local atomic environment, as parametrized by bond distance,
angles, and quantities involving more atoms.

Featuring the coordination number $Z$ in environment-dependent
energy functionals for Silicon, Tersoff \cite{tersoff} proposes $Z$-dependent
bonds, while Bazant et al \cite{bazant} formulate $Z$-dependent 2- and
3-body potentials to better capture the  angular behavior.
Making no assumptions for Boron,
we  investigate this type of description
in its most general form of a $Z$-dependent N-body expansion,
\begin{equation}
\label{eq-gen}
E_i =  \sum_{j}{f(Z_i,r_{ij})} +\sum_{ j \neq
k}{g(Z_i,r_{ij},r_{ik},\theta^{(i)}_{jk})} + ...  
\end{equation}
where $j, k \neq i$ labels neighbors to atom $i$, $r$ the magnitude of
the interatomic distance, and $\theta^{(i)}_{jk}$ the angle formed by
two neighbors $j, k$ and center $i$.

\rem{10 line overview}
In this letter we discuss a structural approach that naturally
 determines such  potentials and allows straight forward
 assessment of the angular forces without assuming an analytic form.
 With an extensive energy database of the so-called ``uniform''
 structures, constructed to have nearly identical bond lengths but
 large angular variation, we sample bonding mechanisms in widely
 different topologies and configurations.  Focusing on 6-coordinated
 ($Z6$) structures, which are relevant to the real phases, we
 determine that their cohesive energies can be sufficiently described
 by $Z6$  potentials of the form
\begin{equation}
\label{eq-Z6}
E_i^{Z6} =  \sum_{j}{f(r_{ij})} +\sum_{j \neq
k}{g(r_{ij},r_{ik},\theta^{(i)}_{jk})} + h(\xi_i)
\end{equation}
where $j, k$ are now restricted to ``nearest'' neighbors of i, and the
 last term is a function of the site asymmetry $\xi_i$ to be defined
 later.

In a ``uniform'' structure \cite{wells}, all sites are geometrically
 equivalent by space-group symmetry operations, so that there is only
 one type of atomic environment occuring throughout the entire
 structure.  We investigate those uniform structures in which all the
 nearest neighbor distances are equal (or nearly equal), so that
 angular effects are more clearly detected.  Our database of over 60
 uniform structures \cite{PRB}, spanning coordination numbers $Z$ from
 3 to 12, and exhibiting large angular variations for each $Z$, serves
 as a set of simple but topologically distinct representatives for
 different regions of configuration space.  In fact, Boron compounds
 adopt certain of these structures as the sub-lattice for the Boron
 atoms \cite{SciAm}.

\begin{figure}
\epsfxsize=3.4in {\epsfbox{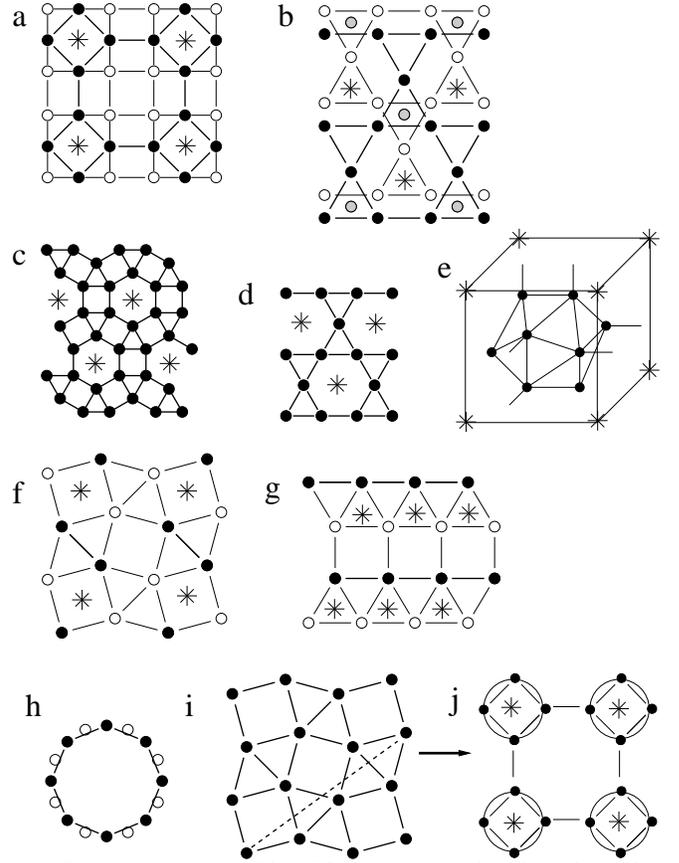}}
\caption{The less familiar $Z6$ structures, with Bravais lattice indicated by
 $\ast$.  For structures in layer form, $\bullet$ are nearest to the
viewer, and $\circ$ furthest.  (a) SC-co ($\bullet \bullet \circ$
stacking), (b) T-lattice, in which the Kagome layer has an (ABC) type
stacking, (c) i-dodec, (d) Kagome, (e) PO-ico, (f) $P\sigma$, (g)
$P\mu$, (h) tube-tri, (i)\&(j) tube-$\sigma$ (see text).}
\label{f-struct}
\end{figure}

$Z6$ coordinations are particularly important in Boron bonding, as
they show up in $80\%$ of the sites in $\beta_{105}$, and $50\%$ in
$\alpha_{12}$.  Therefore we focus now on the uniform $Z6$ structures,
their bonding, and the effective potentials determined from them.
Other coordinations
\rem{, and  analysis of symmetry-breaking relaxations, }
are discussed in \cite{PRB}.  In our list of $Z6$ structures (see
Figure~\ref{f-struct}), besides the simple cubic (SC) and the triangular (tri)
lattices, there are two 3D tilings: the simple cubic array of
cuboctahedra (SC-co) and the diamond array of vertex-joint tetrahedra
(T-lattice) \cite{okeefe}.  From two $Z4$ planar lattices, the 3.4.6.4
(``i-dodec,'' for resembling a set of inter-penetrating dodecagons)
and the 3.6.3.6 (Kagome), we get $Z6$ structures by stacking them with
a separation of unit length (defined as the nearest neighbor
distance.)  There is also the Primitive Orthorhombic packing of
icosahedra (PO-ico) \cite{SL} with inter-icosahedral distance set at
unity.

We have constructed some novel Z6 structures as well.  When we pucker
a planar lattice, and mirror the operation in the next layer, each
vertex gains one additional neighbor from either the layer above or
below.  So from the $Z5$ lattices $3^2.4.3.4$ (called $\sigma$) and
the $3^3.4^2$ (called $\mu$), we get the $P\sigma$ and $P\mu$ $Z6$
structures, (``$P$'' for puckered).  When we roll the triangular lattice
into a single infinite tube with a circumference of 8 edges, we get
the ``tube-tri'' structure \cite{alex}.  Again, by rolling the
$3^2.4.3.4$ ($\sigma$) lattice into a tube with a circumference shown
by the dotted line in Figure~\ref{f-struct}-i, and arranging such
tubes symmetrically in a square array (view along tube axes in
Figure~\ref{f-struct}-j), we get the ``tube-$\sigma$'' structure, in
which each atom has 5 neighbors on its own tube, and one more from a
neighboring tube.

We show in Table~\ref{t-geom} the distribution of ``geometric
quantities'' found in the coordination environments of our $Z6$
structures, which will become helpful information in potential
fitting.  These consist of $N_2$ and $N_3$, the number of neighbors on
the second and third shell, (at distances of $\sqrt{2}$ and
$\sqrt{3}$, respectively), and frequencies of various ``two-neighbor''
angles, (which are formed at a central site and point toward two
nearest neighbors).  We also describe the asymmetry of these atomic
environments, parametrized by
\begin{equation}
\label{e-asym}
\xi =  \frac{| \sum_j {\bf v}_j |}{ (1/Z) \sum_j |{\bf v}_j|}
\end{equation}
where $j$ labels the nearest neighbors, and ${\bf v}_j$  the vector
that points from the central atomic site to  $j$.
The denominator is just the average distance to nearest neighbors.

\rem{LDA Results}

We perform {\it ab-initio} total energy calculation for each structure
in our database \cite{BZ}, in the local density approximation
\cite{teter} with extended norm and hardness conserving
pseudo-potentials \cite{psp}.

\begin{figure}
\epsfxsize=3.6in {\epsfbox{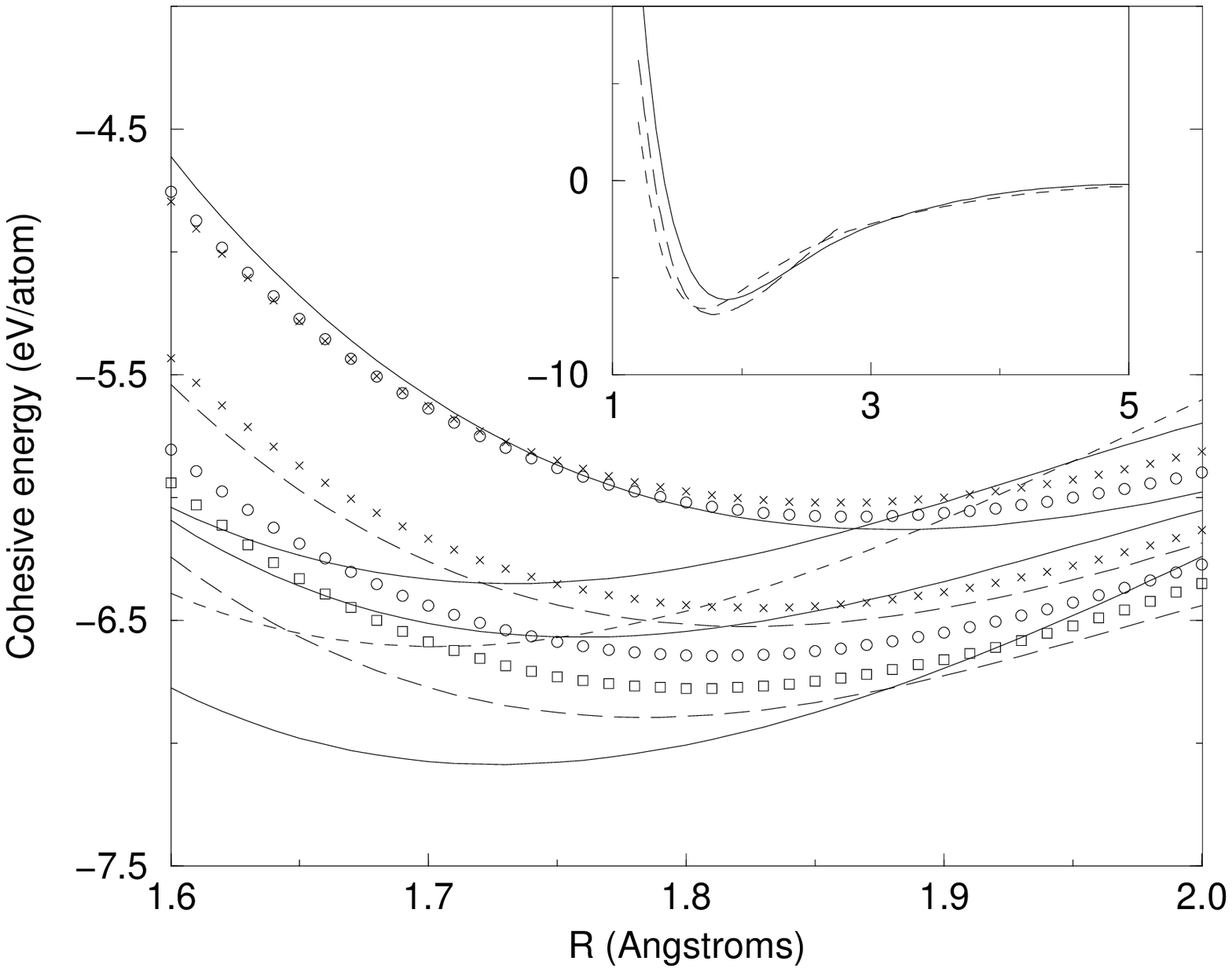}}
\caption {Cohesive energy $E_i(R)$ for $Z6$ structures.  In the inset, the
 three representative structures are SC (line), tri (short-dash), and
 PO-ico (long-dash). The region showing the energy wells is magnified
 in the main figure.  The structures are, in order of lowest to
 highest energy at $R = 1.8\AA$, ideal-$\alpha_{12}$ (line), PO-ico
 (long-dash), SC-co ($\Box$), tube-$\sigma$ ($\circ$), T-lattice
 (line), P-$\mu$ (long-dash), tri (short-dash), P-$\sigma$ ($\times$),
 tube-tri (line), SC (line), i-dodec ($\circ$), and Kagome
 ($\times$).}
\label{f-Er}
\end{figure}

For each $Z6$ structure, the edge length ``$R$'' representing the
 nearest neighbor distance may be varied.  In Figure ~\ref{f-Er}, we
 present the cohesive energy (per atom) of the structure as a function
 of $R$ in a range of physical interest.  In the inset, where $R$
 scales from $1 - 5 \AA$, we show the resulting single-well curves for
 three representative structures, SC, tri, and PO-ico
\cite{ftnt-enecurve}. 
The energy curves of i-dodec and Kagome fall closely onto the SC
curve, and similarly the T-lattice onto the tri, and SC-co onto the
PO-ico.  Beyond $R = 3 \AA$, all curves converge, and saturate toward
the non-bonding limit.

In the main figure, we enlarge the $[1.6, 2.0] \AA$ region, (within which
 bond lengths from different structures all fall \cite{alex}),  to clarify
the difference among the wells of the energy curves.  Our lowest
energy $Z6$ structure is the PO-ico, a topologically different
icosahedral packing.  Within the next low energy structure, SC-co, is
the cuboctahedron that has been compared to a distorted icosahedron
using molecular orbitals theory \cite{bullett}.  For comparison, we
include the lowest energy curve of the ``ideal-$\alpha_{12}$'' phase,
topologically same as $\alpha_{12}$ but with all nearest neighbor
distances set equal.

To construct an effective theory, it is natural to start with a
typical single-well 2-body potential, with nearest neighbors close to
the minimum of the well, and further neighbors contributing at the
tail of the well.  This will not work for our $Z6$ energy curves.  The
large variation of $N_2$ and $N_3$ in Table~\ref{t-geom} implies that
the energy contribution from the further neighbors will lead to a
large energy variation, on the order of the well-depth, among the
structures.  This contradicts the small energy differences among our
energy curves in Figure~\ref{f-Er}.  Indeed, numerical fittings with
several single-well forms  all substantiate this argument.  We conclude
 that the 2-body potential has to be independent of
$N_2$ and $N_3$, i.e. it must be essentially truncated after
nearest-neighbor range.
\rem{This offers an alternative argument to the same result as
obtained by inversion methods \cite{bazant}.}

Since we have shown the important result that the first term in
Eq.~\ref{eq-gen} sums only 
the nearest neighbor $j$, our $Z6$ structures must have basically the
same 2-body contribution of $6f(R)$ \cite{ftnt-6fr}.  The energy
differences among these structures would have to come from
higher-order potentials.  We choose the simple cubic energy curve to
serve as the arbitrary reference in taking energy differences, and to
give the function $f(x)$.  We propose a 3-body potential that depends
only on the two-neighbor angles.  We also include a term $h(\xi)$
(compare Eq.~\ref{eq-Z6}) linear in $\xi$ to test the degree of energy
dependence on the asymmetry,  which involves all $Z$ neighbors.
Allowing for dependence on the scale $R$, the cohesive
energies (per atom) would lead to a set of linear relationships,
\begin{equation}
\label{e-linrel}
\Delta E_m(R) = \sum_{a   }{ N_m(\theta_a) g(R,R,\theta_a)} + c(R) {\xi}_m
\end{equation}
where $\Delta E_m$ is the energy difference between the $m$-th
structure and the reference, $a$ labels the two-neighbor angles, $N_m$
the number function per atom, \rem{for the $m$-th structure} and $c$
the linear coefficient of $\xi$ that depends on $R$.

A major problem in determining the functional form of the angular
potential $g(R,R,\theta)$ is that the frequency of a particular angle
$\theta_a$ cannot be varied independently of $\theta_{a'}$ to display
its individual effect on the cohesive energy.  Normally this problem
is avoided by fitting to an assumed form \cite{bazant}, but for Boron,
where angular behavior is still being explored, we have no analytic
guidance as to the form of this potential.  However, the small number
of distinct angles $\theta_a$ appearing among the structures in
Table~\ref{t-geom} helps to resolve this.

\begin{figure}
\epsfxsize=3.6in {\epsfbox{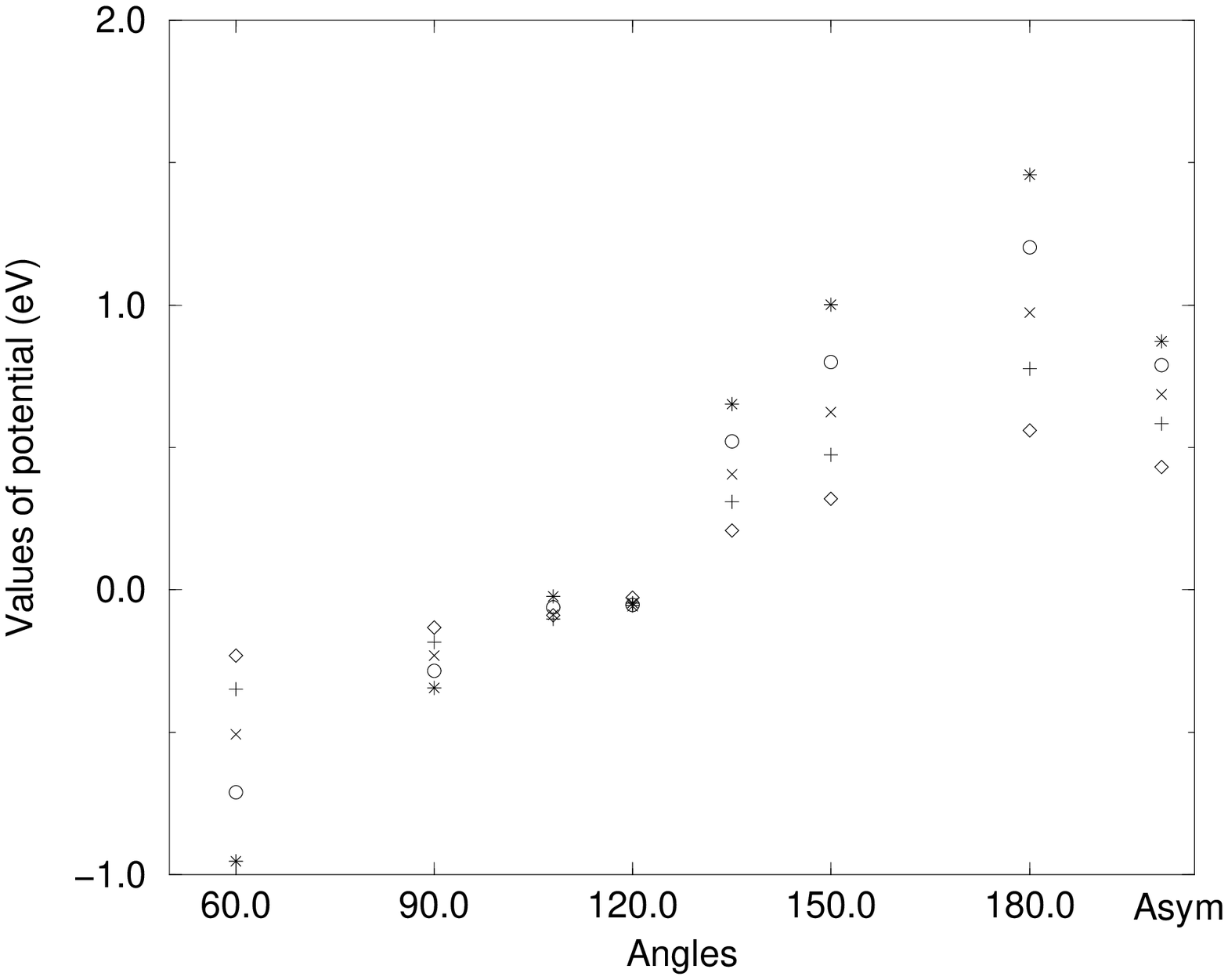}}

\caption {Angular and asymmetry potentials.
The angles take on values from \deg{60} to \deg{180} .  Data in the
``Asym'' column signifies the coefficient $c(R)$ in the asymmetry
potential.  The $g(R,R,\theta)$ and $c(R)$ are illustrated by symbols
$\{ *, \circ, \times, +, \diamond \}$, corresponding to $R= \{ 1.6,
1.7, 1.8, 1.9, 2.0 \} \AA$.}
\label{f-pot}
\end{figure}

Since all nearest neighbor lengths are nearly equal, we can separate
the $R$-dependence from the angular effects.  The low number of
distinct angles that appear in our $Z6$ structures allows us to
determine the shapes of the angular potential without further
assumptions.  With the eleven $Z6$ structures and the data from
Table~\ref{t-geom}, we can solve numerically, in the least-squares
sense, for $g(R,R,\theta_a)$ and $c(R)$, not assuming any functional
form.  Doing this at different $R$ gives the radial dependence of $g$
and $c$.  The resulting fit, shown in Figure~\ref{f-pot}, suggests
that the angular potential is linearly monotonically increasing.  The
$R$-dependence for each angle and $c(R)$ shows rapid decay.  Similar
potentials are found for Z5 structures \cite{PRB}.

Although the angular potential is obtained from the correlation of the
number of angles with total energy, and not from measuring the direct
effects each angle has on the site energy, this result is by no means
trivial: for the potential value at each of the 7 angles is an
independent parameter in the fit, yet the resulting $g(R,R,\theta)$
looks smooth as a function of $(R,\theta)$.  Furthermore, we find that
a separable form, i.e. $g(r_1,r_2,\theta) = \hat g(r_1)\hat
g(r_2)A(\theta)$, is sufficiently determined, and a good
approximation.

When we apply our potentials $g$ and $c$ to the $Z6$
``inverted-umbrella'' environment \cite{shirai} as found in the
ideal-$\alpha_{12}$ structure, its site energy is 1.0 eV lower than
that of the second best environment (PO-ico) among our structures.  In
fact, a Monte Carlo search among $Z6$ environments having all
two-neighbor angles greater than \deg{60} found no lower energy
configurations.  Since the effective potentials were constructed from
environments all higher in energy, it is encouraging that they predict
the realistic ``inverted-umbrella'' to be lowest in energy.
Furthermore, our potential disfavoring \deg{180} { }is in agreement
with the tendency for Boron clusters to buckle in a hexagonally
coordinated environment \cite{alex}.

The angular and asymmetry potentials constructed so far can account
for most of the 1-2 eV/atom energy variations among the $Z6$
structures, with an average error of 0.14 eV/atom in the bonding
range.  The energy ordering of our $Z6$ structures are mostly retained
by the potentials.  Although we still need to treat other Z's, and
test cases when not all neighbors are at the same distance, we see a
promising start in the local potential description for the general
Boron system, without many-body terms involving, e.g. all the atoms in
an icosahedron.

\rem{acknowledgement}
We thank M. Teter, D. Allen,  J. Charlesworth for providing 
code and support in the LDA calculation, A. Quandt, M. Sadd for  comments,
\rem{E. Kaxiras,}  K. Shirai, P. Kroll and R. Hoffman for discussion.
This work is supported by DOE grant DE-FG02-89ER45405.

\vfill\eject

%%%%%%%%%%%%%%%%%%%%%%%

%\section{Tables and Figures}

\begin{table}
\caption {Local environments of   $Z6$ structures.}
\label{t-geom}
\begin{tabular}{lcccccccccc}
Structure  &  $N_2$  &  $N_3$  &  \multicolumn{7}{c}{$N(\theta_a)$}& $\xi$\\
  	   & 		& 	&

$\deg{60}$ & $\deg{90}$ & $\deg{108}$ & $\deg{120}$ & $\deg{135}$ &
$\deg{150}$ & $\deg{180}$ & \\
\tableline
SC	& 12& 8& 0& 12& 0& 0& 0& 0& 3& 0\\
 tri	& 0&  6& 6& 0&  0& 6& 0& 0& 3& 0\\
SC-co	& 7&  6& 2& 7&  0& 2& 4& 0& 0& 0.586\\
T-lattice & 0& 12& 6& 0&  0& 6& 0& 0& 3& 0\\
i-dodec	& 10& 6& 1& 10& 0& 1& 0& 2& 1& 0.732\\
Kagome	& 8&  4& 2& 8&  0& 2& 0& 0& 3& 0\\
PO-ico	& 1& 10\tablenote{Neighbors between the second and third
 shells, occuring at various 
distances,  are binned in $N_3$.}
& 5& 1&  7& 0& 0& 2\tablenote{These bins have tolerance of 
0.14 unit distance,  or $\deg{6}$ in angle.}
& 0& 1.854\\
$P\sigma$ \tablenote{Nearest neighbors not all at unit distance
\cite{ftnt-6fr}.}  & 4$\rm {}^b$& 12$\rm {}^a$& 3& 3$\rm {}^b$& 4$\rm
{}^b$& 1$\rm {}^b$& 2$\rm {}^b$& 2$\rm {}^b$& 0& 0.263\\ 
$P\mu$ & 5$\rm {}^b$& 10$\rm {}^a$& 3& 4& 3$\rm {}^b$& 2& 2$\rm
{}^b$& 0& 1& 0.771\\ 
tube-tri & 0& 6$\rm {}^a$& 6& 0& 0& 6$\rm {}^b$&
1& 0& 2\tablenote{This is at \deg{169} . } & 1.163\\ 
tube-$\sigma$ $\rm {}^c$& 5$\rm {}^b$& 7$\rm {}^a$& 3& 5$\rm {}^b$&
2$\rm {}^b$& 2& 0& 3$\rm {}^b$& 0& 0.936\\
\end{tabular}
\end{table}

\begin{references}
\bibitem{BZ}   We use all
 planewaves with kinetic energy up to 54.5 Ry.  The Brillouin zone is
sampled with a k-point density of at least $(16.3\AA^{-1})^3$.
Band structure energies are converged to within $10^{-6}$ Ry in each
calculation. The convergence with respect to k-point density shows a
precision of 0.03 eV for energy comparison among different structures.
\bibitem{ftnt-enecurve}
For each structure, with unit length defined as $1.631 \AA$, the
shortest bond length found in the $\alpha_{12}$ phase, we sampled 10
energy data points at increment of 2\% dilation within the single-well
region, and then at total dilations of 0.7, 0.8, 0.9, 1.5, 2.0, 3.0
and 5.0.  We present here the spline fitted curve.
\bibitem{ftnt-6fr}
$P\sigma$ has 4 nearest neighbors at 1.03108 R, and tube-$\sigma$ has 2
at 1.0072 R  and 1 at 1.0256 R.  These corrections are accounted for
in the 2-body energies.
\end{references}
\end{document}